\documentstyle[12pt]{article}

\input epsf
\newwrite\ffile\global\newcount\figno \global\figno=1

\def\writedef#1{}

\def\figin{\epsfcheck\figin}\def\figins{\epsfcheck\figins}
\def\epsfcheck{\ifx\epsfbox\UnDeFiNeD
\message{(NO epsf.tex, FIGURES WILL BE IGNORED)}
\gdef\figin##1{\vskip2in}\gdef\figins##1{\hskip.5in}
\else\message{(FIGURES WILL BE INCLUDED)}%
\gdef\figin##1{##1}\gdef\figins##1{##1}\fi}

\def\figinsert{}
\def\ifig#1#2#3{\xdef#1{fig.~\the\figno}
\writedef{#1\leftbracket fig.\noexpand~\the\figno}%
\figinsert\figin{\centerline{#3}}\medskip\centerline{\vbox{\baselineskip12pt
\advance\hsize by -1truein\center\footnotesize{  Fig.~\the\figno.} #2}}
\bigskip\endinsert\global\advance\figno by1}
\def\endinsert{}

\begin{document}

\title{\large{\bf
Controlled Soft Breaking of N=1 SQCD}}

\author{
\\
Nick Evans\thanks{nevans@physics.bu.edu}
\\ \\ Department of Physics, Boston University, Boston, MA 02215 \\ \\ \\
Stephen D.H.~Hsu\thanks{hsu@hsunext.physics.yale.edu},
Myckola Schwetz\thanks{ms@genesis2.physics.yale.edu}
\\ \\ Department of Physics, Yale University, New Haven, CT 06520-8120 \\
}

\date{March 28, 1997}

\maketitle

\begin{picture}(0,0)(0,0)
\put(350,350){BUHEP-97-10}
\put(350,335){YCTP-P2-97}
\end{picture}
\vspace{-24pt}

\begin{abstract}
We discuss the introduction of soft breaking terms into the exact solutions of
N=1 SQCD using a spurion analysis. The spurion symmetries are not
sufficient to determine the behavior of models in which
squark or gaugino masses alone are introduced. However, a
controlled approximation
is obtained in some cases if a supersymmetric mass is first introduced for
the matter fields. We present low-energy solutions for two  models with
perturbing soft breaking terms, one with a gaugino mass and one with
squark mixing. These models have non-trivial theta
angle dependence and exhibit phase transitions at non-zero theta angle
analogous to those found in the chiral Lagrangian description of QCD.
\end{abstract}

\newpage
\section{Introduction}

The exact solutions \cite{seiberg1, seiberg2, seiberg3}
for the IR Wilsonian effective theory of N=1
supersymmetric QCD (SQCD) reveal some surprising dynamical effects.
Most striking are the occurence of massless composite bound states (or
solitons) in the strong coupling regime. It is intriguing
whether these massless states could smoothly map to states important to
the dynamics of non-supersymmetric gauge theories.
It is highly implausible that the massless composite
fermions of SQCD can  survive in the QCD limit.  The
lattice arguments of Weingarten \cite{Wein} imply that
any composite states in QCD  must be heavier than the pions.
Nevertheless, it is possible, for example,
that the scalar ``dual quark'' solitons might
survive in some form and be involved in some ``dual magnetic''
description of  confinement in QCD.

Soft breaking terms, such as squark and gaugino masses,  may be
introduced to the SQCD theories as spurion fields with non-zero
F-component vevs that explicitly break supersymmetry \cite{soft1}.
The symmetries of the enlarged spurion model
constrain how they may appear in the low energy Wilsonian
theory. In general these constraints are not however sufficient to
determine the low energy theory since ``Kahler Potential'' terms may be
constructed that are invariant to all symmetries and are hence
unknown \cite{soft2}. For the cases where a squark and/or gaugino mass are the
sole
supersymmetry and chiral symmetry breaking parameters these Kahler terms
dominate the behaviour of the potential. Some speculations as to the
behaviour of these theories were made in Refs \cite{peskin,hoker}.

In this paper we discuss these difficulties and investigate some cases in which
the effects of the soft breakings {\it can} be controlled.
We start with a model with supersymmetry preserving quark/squark masses,
and then break supersymmetry with squark and
gaugino masses resulting from spurions that occur linearly in the
superpotential. It can be shown that any possible Kahler corrections are higher
order in the soft breakings, and thus control may be retained over the
low-energy theory. The analysis is
similar to that performed on the N=2 SQCD solutions in Ref
\cite{soft2}.  The
derivative (low-energy) expansion performed to obtain
the solutions of SQCD restricts the
solutions of the softly broken models to the regime where the soft
breakings are  small relative to the strong interaction scale.  At first
sight the resulting models appear to behave almost identically to their
supersymmetric counterparts but, as for the N=2 solutions \cite{soft3},
the models
have the additional new feature of displaying $\theta$ angle
dependence. The softly broken models distinguish the $N_c$ vacua of the
SQCD models and as $\theta$ is changed these vacua interchange at first
order phase transitions. We contrast this behaviour with that of the QCD
chiral Lagrangian.

\section{N=1 SQCD}

We begin from the N=1 $SU(N_c)$ SQCD theories with $N_f$ flavors
described by the UV Lagrangian
\begin{equation}
{\cal L} = K^\dagger K (Q^\dagger_i Q_i + \tilde{Q}^\dagger_i
\tilde{Q}_i)|_D + {1 \over 8 \pi}
Im \tau W^\alpha W^\alpha|_F + 2 Re\, m_{ij} Q_i \tilde{Q}_j|_F
\end{equation}
where $Q$ and $\tilde{Q}$ are the standard chiral matter superfields and
$W^\alpha$ the gauge superfield. The coupling $K$
determines the kinetic normalization of the matter fields. The
gauge coupling $\tau = \theta/2 \pi + i 4 \pi/g^2$  defines
a dynamical scale of SQCD:
$\Lambda^{b_0} = \mu^{b_0} exp( 2\pi i \tau)$,
with $b_0 = 3 N_c - N_f$ the one loop coefficient of the SQCD
$\beta$-function.
And, finally, $m$ is  a supersymmetric mass term for
the matter fields. We may raise these couplings to the status of spurion
chiral superfields which are then frozen with scalar
component vevs. The SQCD theory
without a mass term has the symmetries
\begin{equation}
\begin{tabular}{ccccc}
&$SU(N_f)$ & $SU(N_f)$ & $U(1)_B$ & $U(1)_R$\\
$Q$ & $N_f$ & 1 & 1 & ${N_f - N_c \over N_f}$\\
$\tilde{Q}$ &1& $\bar{N}_f$ & -1 & ${N_f - N_c \over N_f}$\\
$W^\alpha$ & 1 & 1 & 0 & 1\end{tabular}
\end{equation}
The mass term breaks the chiral symmetries to the vector symmetry. The
classical $U(1)_A$ symmetry on the matter fields is anomalous and,
if there is a massless quark, may be used to rotate away the theta
angle. In the massive theory the flavor symmetries may be used to
rotate $m_{ij}$ to diagonal form and the anomalous $U(1)_R$ symmetry
under which the $Q$s have charge $+1$ may be used to rotate $\theta$ on
to the massless gaugino. Including
the spurion fields the non-anomalous $U(1)_R$ symmetry charges are
\begin{equation}\label{sym}
\begin{tabular}{cccccc}
$W$ & $Q$ & $\tilde{Q}$ & $\tau$ & $m$ & $K$ \\
1 & ${N_f - N_c \over N_f}$ & ${N_f - N_c \over N_f}$ & 0 & ${2N_c \over
  N_f}$ & {arbitrary} \end{tabular}
\end{equation}
The anomalous symmetries may be restored to the status of symmetries of
the model if we also allow the spurions to transform. The appropriate
charges are
\begin{equation}
\begin{tabular}{ccccccc}
&$W$ & $Q$ & $\tilde{Q}$ & $\Lambda^{b_0}$ & $m$ & $K$ \\
$U(1)_R$ & 1 & 0  & $ 0 $ & $2(N_c-N_f)$ &  2  & arbitrary\\
$U(1)_A$ & 0 & 1  & 1       & $2N_f$             & -2 &  arbitrary

\end{tabular}
\end{equation}
The $m_{ij}$ spurions also transform under the
chiral flavor group.

The solutions of the models are $N_f$ dependent. For $N_f < N_c$ the
low energy superpotential is exactly determined by the symmetries and
the theory has a run away vacuum \cite{seiberg1}. For $N_f = N_c$ the low
energy theory
is in terms of meson and baryon fields
\begin{eqnarray}
M_{ij} & = & Q_i \tilde{Q}_j \nonumber\\
b^{[i_1,...,i_N]} & = & Q^{i_1} ... Q^{i_{N_c}}\\
\tilde{b}^{[i_1,...,i_N]} & = & \tilde{Q}^{i_1} ... \tilde{Q}^{i_{N_c}}
\nonumber
\end{eqnarray}
subject to the constraint $det M
+ b \tilde{b} = \Lambda^{2N_f}$ \cite{seiberg2}. For $N_F = N_c+1$
the theory is again described
by baryon and meson fields with the classical moduli space unchanged
\cite{seiberg2}.

When $N_c+1 < N_f < 3N_c$ the theory has
an alternative description of the low energy physics in terms
of a dual magnetic theory with an $SU(N_f-N_c)$ gauge group, $N_f$
flavors of dual quarks, $q$ and $\tilde{q}$, and $N_f^2$ meson fields,
 $M_{ij}$ \cite{seiberg3}.
The dual theory has the
additional superpotential term $M_{ij}q_i \tilde{q}_j$. Generally one of the
two
duals is strongly coupled whilst the other is weakly coupled (the
electric theory is weakly coupled for $N_f \sim 3N_c$, the magnetic
theory when $N_f \sim N_f+2$). In the strongly coupled variables
the low energy Wilsonian effective
theory is a complicated theory with all higher dimensional terms in the
superfields equally important (since the IR theory is in a conformal
regime the scale $\Lambda$ at which the theory entered the conformal
regime is not available to suppress higher dimension terms and similarly
the gauge coupling is order one and may not suppress these
operators). The weakly interacting theory however, has a very simple
Wilsonian effective theory of the canonical bare form. According to the
duality conjecture these two effective theories must describe the same
physics and therefore there is presumably a (complicated) mapping
between the electric and magnetic variables in the IR.

\section{Soft Supersymmetry Breaking}

Soft breaking interactions terms which explicitly break supersymmetry
may be included in the UV theory by allowing the spurions to acquire
non-zero $F$-components.(These are the terms  that can be induced by
spontaneous supersymmetry breaking and hence may be included
perturbatively while  inducing only logarithmic divergences in the
theory as a remnant of the supersymmetric non-renormalization theorems
\cite{soft1}).
We will consider
three such breaking terms, a squark mass ($F_K \neq 0$), a gaugino
mass ($F_\tau \neq 0$) and a squark mass mixing ($F_m \neq 0$).

The dependence of the IR effective
theory on the spurion fields is
determined in the N=1 limit by the
dependence on their scalar components, the couplings and masses.
The exact solutions of Seiberg, however, do not provide
sufficient information to take the soft breakings to infinity limit and
obtain results for models with completely decoupled superpartners since
the solutions are only low energy derivative expansions. Higher
dimension terms are suppressed by the strong coupling scale $\Lambda$
and hence in the non-supersymmetric theories there are unknown soft
breaking terms of higher order in $F_S / \Lambda^2$.

A second problem is that squark masses are only generated through the
Kahler potential (the spurion $F_m$ generates a squark mass mixing but it
is unbounded without additional contributions to the masses from the
Kahler sector) via such terms as $|F_S|^2 |Q|^2$ with $S$ a general
spurion.  There are no symmetry constraints on these terms so we do not
know whether they occur in the low energy theory or if they do, their
sign. We note that the sign of these terms relative to the sign of the
equivalent terms in the UV theory is crucial.
As a particular example consider theories close to $N_f = 3N_c$ where the
electric theory has a very weak IR fixed point and the magnetic theory a
strongly coupled IR fixed point. We are interested in what happens when
we introduce squark and gaugino masses in the UV magnetic theory. We can
consider the case where these soft breakings are small relative to the
scale $\Lambda$ at which the theory enters it's strongly interacting
conformal phase. We
expect a conformal phase down to the soft breaking scale but can we say
anything about the theory below that scale? The dual squarks in the
weakly coupled IR description only
acquire masses from $F_\tau$ and $F_K$ from the Kahler terms. For
infinitesimal soft breakings we do not expect the weakly coupled  nature
of the dual theory at the breaking scale to be disturbed. If these
masses are positive (as investigated in Ref\cite{peskin}) then below the
soft breaking scale the theory behaves like QCD and presumably confines
and breaks chiral symmetries at an exponentially small scale relative to
the soft breaking masses. Alternatively if the masses are negative (as
investigated in Ref\cite{hoker}) then the magnetic gauge group is
higgsed with the possible interpretation in the electric variables of a
dual Meissner type effect. The spurion symmetry arguments are not
sufficient to distinguish between these possibilities.

It should be remarked that there {\it is} a strongly coupled magnetic
theory that corresponds to the introduction of any soft breaking terms
in the electric theory. This is true since we can use the
mapping of
electric to magnetic field variables from the SQCD theory
(which is not known explicitly, but exists in principle)
to
write the soft breaking terms of the simple weakly interacting theory  in terms
of the
strongly interacting variables in the IR. The result will be a
complicated mess of relevant higher dimension operators in the strongly 
interacting
theory. The subtlety is that if we now run the renormalization group back to 
the UV in the
magnetic variables we will, very likely, never recover a weakly
interacting theory. At each step to recover the effective theory at the
lower  scale we must add important higher dimension terms. The problem
is therefore to identify which soft breaking terms in the IR electric
description correspond to canonical soft breaking terms in the UV
magnetic theory.

In the next section we shall resolve this problem for the $F_\tau$ and
$F_m$ cases after including a supersymmetric mass that determines the
squark masses at order $F^0$. Then for small soft breakings relative to
$m$ (and $\Lambda$) exact solutions may be obtained.

\section{Controlled N=0 Theories}

To obtain  solutions to softly broken N=1 SQCD theories,
we begin by including a supersymmetric mass for the matter fields.
The resulting theories
have a mass gap on the scale $m$ and the induced meson $M_{ij}= Q^i
\tilde{Q}_j$ vev is determined independently of $N_f$ by holomorphy
\begin{equation}\label{Slimit}
M_{ij} = \Lambda^{{3N_c - N_f \over N_c}}  (detm)^{1/N_c}\left( {1 \over
  m} \right) _{ij} = |M_{ij}| e^{i\alpha}~~~.
\end{equation}
The resulting supersymmetric theories have $N_c$ distinct vacua
corresponding to the $N_c$th roots of unity, $\alpha = 2n\pi/N_c$
(as predicted by the Witten
index). Note that for the theories
with magnetic duals putting masses in for all flavors breaks the dual
gauge group completely. For simplicity henceforth we shall take $m_{ij}$
to be proportional to the identity matrix; in this basis $\langle
M_{ij} \rangle$ is also proportional to the identity matrix.

These massive theories may be softly broken in a controlled fashion.
If the spurion generating the soft breaking enters
the superpotential linearly then we may obtain desirable results when that
spurion's F-component $F \ll m \ll \Lambda$. Any D-term contributions to
the scalar potential take the form $F_X^\dagger F_Y$ with $X$ and $Y$
standing for generic fields or spurions. In the supersymmetric limit all
F-components are zero and will grow as the vacuum expectation value of the
soft breaking spurion.
These Kahler terms are therefore higher order in
the soft breaking parameter than the linear term from the
superpotential. The unknown corrections to the squark masses in the
theory are subleading to the masses generated by the supersymmetric mass
term and hence we may determine the potential minima at lowest order.

\subsection{Squark Mass Mixing}

The first model we  consider includes  the bare squark mixing term
\begin{equation}
Re(F_{m \, ij} ~Q_i \tilde{Q}_j)
\end{equation}
which is generated from the superpotential. Again for simplicity we will
take $F_{mij}$ to be diagonal with degenerate eigenvalues in the basis
in which $m_{ij}$ is diagonal.
The form of the effective
theory is governed by the symmetries in (\ref{sym}) which determine that the
superpotential of the theory is not renormalized. The soft breaking term
is therefore also not renormalized and
is the leading term in an expansion in
$m/\Lambda$. For $F_m \ll m \ll \Lambda$ we find that there are the
$N_c$ minima of the SQCD theory given by the values of $M_{ij}$ in
(\ref{Slimit})
and distinguished by their contribution to the potential
\begin{eqnarray}\label{Fmpotential}
-Re Tr[ F_{m} M_{ij}]& = & - N_f |F_m| |M|
\cos([\theta_0 + (N_f -  N_c) \theta_m +
N_c  \theta_f + 2n \pi]/N_c])\\
& = & - N_f |F_m| |M| \cos([\theta_{phys}+ 2n \pi]/N_c)~~~. \nonumber
\end{eqnarray}
Freezing the spurion $F_m$ explicitly breaks $U(1)_R$ and introduces
dependence on the $\theta$ angle.
$\theta_{phys}$ is the correct combination of phases on $m$, $F_m$ and the
bare $\theta$ angle. To see this in the bare Lagrangian we
may use the anomalous $U(1)_A$ symmetry
to rotate any
phases on $F_m$ onto $m$ and into the $\theta F \tilde{F}$ term.
Then using the anomalous $U(1)_R$ symmetry under which $Q_i$
transforms with charge 0 we may rotate the resulting phase on $m$ into
the $\theta$ angle as well. The resulting $\theta$ angle is the physical
$\theta$ angle in which the physics is $2 \pi$ periodic:
\begin{equation}
\label{phth}
\theta_{phys} = \theta_0 + (N_f - N_c) \theta_m + N_c \theta_f
\end{equation}
We can also understand the form of (\ref{phth}) as follows.
Once the $U(1)_R$ symmetry
is explicitly broken by $f_m$ a gaugino mass is generated by radiative
effects. We can think of $\theta_{phys}$ as generated by the effective
phases on the quark and gaugino masses. The gaugino mass is generated
by a perturbative graph with a quark-squark loop. The result is of the
form $F_m / m$, leading to an effective phase which is $\theta_f -
\theta_m$.The effective gaugino phase then
appears in (\ref{phth})
with an anomaly factor from $C_2 (R)$ of $N_c$ rather than $N_f$.
The equivalent effective superpotential term is of the form
\begin{equation}
ln [ m ]~ WW~ \vert_F,
\end{equation}
which yields another contribution to the potential when the
gauginos condense. Using the Konishi anomaly \cite{KA},
one can see that
this term has the same form as (\ref{Fmpotential}).

The resulting potential (\ref{Fmpotential})
distinguishes the $N_c$ vacua. For $\theta_{phys} = 0$ the $n=0$ vacua
is the true minima.

\vspace{.4cm}

$\left. \right.$  \hspace{-0.4in}\ifig\prtbdiag{}
{\epsfxsize 12truecm\epsfbox{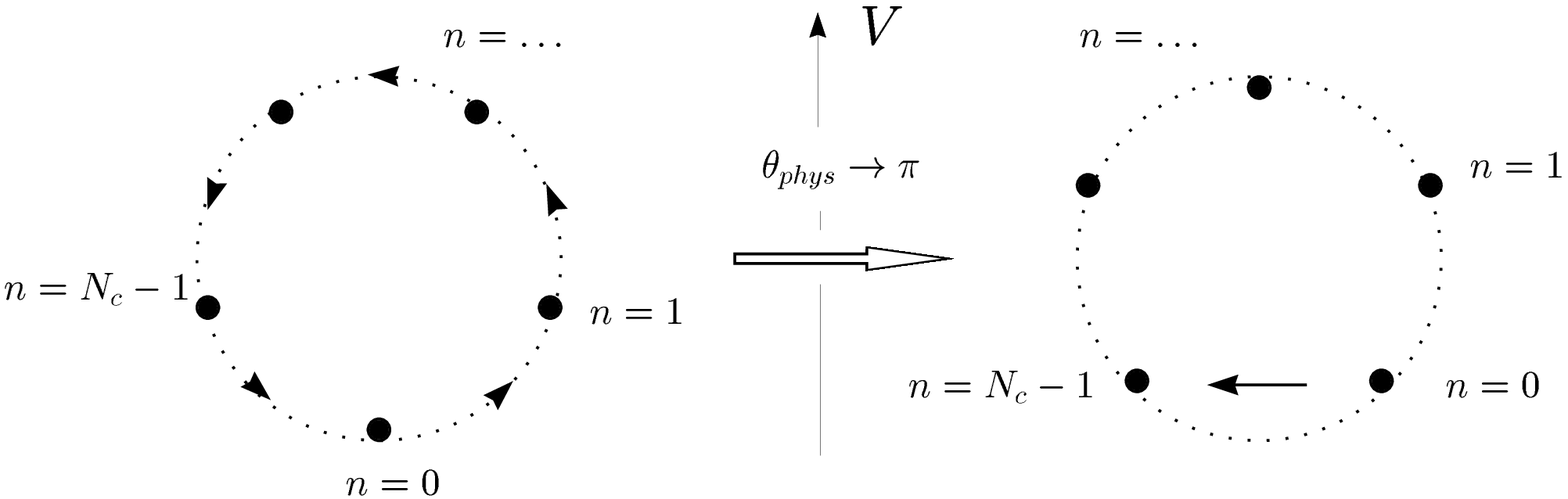}} \vspace{-1.7cm}
\begin{center}
Fig.1\,: First order phase transition as $\theta_{phys}$ is
varied from 0 to $\pi$.
\end{center}

As $\theta_{phys}$ passes through $\pi$ the $n=0,N_c -1$
vacua become degenerate and there is a first order phase transition. Then
as $\theta_{phys}$ moves through (odd)$\pi$ there are subsequent first order
phase transitions at which the SQCD minima interchange.

\subsection{Gaugino Mass}

In the UV theory we may induce a  gaugino mass through a non zero
F-component of the gauge coupling $\tau$
\begin{equation}
{1 \over 8 \pi} Im [ F_{\tau} \lambda \lambda]
\end{equation}
In the IR theory $\tau$ enters through the strong interaction scale
$\Lambda$ which again occurs linearly in the superpotential of the
theory. Taking $F_{\tau} \ll m \ll \Lambda$ we again may determine
the vacuum structure. The IR superpotential terms compatible with the
symmetries of the theory involving $\Lambda$ are
\begin{equation}
Re[ m M_{ij} + ({\rm det} M_{ij})^ { 1/(N_f - N_c) } \Lambda^{(3N_c-N_f)
  / (N_c-N_f)}]
\end{equation}
where the final term results from non-perturbative effects in the broken gauge
group. At lowest order in perturbation theory the vev of $M_{ij}$ is
given by (\ref{Slimit}) which also contains $\Lambda$ and hence has a
non-zero F-component. Including $F_\tau$ and performing the superspace
integral we obtain up to a coefficient the following corrections
to the potential that
break the degeneracy between the $N_c$ SQCD vacua
\begin{eqnarray}
\label{gpot}
\Delta V & = & - Re\left[ m^{N_f/N_c} i F_\tau \Lambda^{(3N_c-N_f)/
                N_c}\right]\\
& = \nonumber & - \left|m^{N_f/N_c} F_\tau \Lambda^{(3N_c-N_f)/
                N_c}\right| \cos[ ~ \theta_{phys}/N_c ~+~ \alpha ~]
\end{eqnarray}
where again $\alpha$ are the $N_c$th roots of unity and
$\theta_{phys}$ is the physical theta angle in which the
physics must be $2 \pi$ periodic. It may be obtained by again making rotations
with the anomalous $U(1)_A$ and $U(1)_R$ symmetries
\begin{equation}
\theta_{phys} ~=~ \theta_0 ~+~ N_c ( \theta_{F_\tau} + 
\pi /2) ~+~ N_f \theta_m
\end{equation}
The factor of $\pi/2$ occurs as a 
result of the discrepancy between the 
phase of $F_\tau$ and that of the canonical definition of the
gaugino mass.
There is also an additional contribution to the vacuum energy
arising from the gaugino condensate. Using the Konishi
anomaly \cite{KA}, we see that it has the same form as
(\ref{gpot}).
The supersymmetry breaking contributions again break the degeneracy
between the $N_c$ supersymmetric vacua. There are again phase transitions
as $\theta_{phys}$ is varied,  occurring
at $\theta_{phys} ~=~ $(odd)$\pi$.

\section{Discussion}

We have investigated some examples where controlled, low-energy descriptions of
softly broken massive SQCD may be obtained, despite the lack of supersymmetry.
The models we studied are obtained by the inclusion of
soft breaking masses from spurions occuring linearly in the
superpotential. Examples of such soft
breaking terms are gaugino masses and squark mass mixings.The soft
breaking corrections to the potential distinguish between the $N_c$
vacua of SQCD at a generic value of theta angle. At the special values of
$\theta_{phys} = $(odd)$\pi$ there are first order phase transitions at which
two of the $N_c$ vacua interchange.

This behavior can be compared
with the theta angle dependence of the QCD chiral Lagrangian \cite{chiral}
for which there are $N_f$ distinct vacua which interchange through first
order phase transitions at $\theta =$(odd)$\pi$. This difference in the
number of vacua between the softly broken theories and QCD would prohibit us
from seeing any sign of a smooth transition between the two theories
(one might hope that the $M_{ij}$ vev might smoothly map to the quark
condensates of QCD for example) even if we were able to begin to
take the squark and gaugino masses towards infinity. There is however
one conclusion for QCD that we can tentatively draw from this
analysis. In these models the form of the confined effective
theory changes smoothly with the theta angle and there is no sign of a
break down of confinement as suggested in \cite{schierholz}. This lends some
support to the assumption \cite{chiral} that the chiral Lagrangian remains
the correct discription of QCD in the IR even at non-zero theta.

\vspace{3cm}

\noindent {\Large \bf Acknowledgements}

NE would like to thank R. Sundrum for useful discussions. This work 
was supported by DOE contracts DE-AC02-ERU3075 and DE-FG02-91ER40676.

\newpage

\end{document}